\documentclass{elsart5p}

\usepackage{graphicx}

\usepackage{amssymb}

\usepackage{lineno}

\begin{document}

\begin{frontmatter}



\title{RICAP-07: Summary Comments}
\author{T.K. Gaisser\corauthref{cor1}}
\address[cor1]{Bartol Research Institute and Dept. Physics \& Astronomy, 
University of Delaware, Newark DE, 19716, U.S.A.}
\ead{gaisser@bartol.udel.edu}


\author{}

\address{}

\begin{abstract}
The Roma International Conference on Astroparticle Physics
covered gamma-ray astronomy, air shower experiments and neutrino astronomy
on three successive days.  I organize my brief summary 
comments into four topics that cut across these three techniques.
They are detector calibration, galactic sources, extra-galactic sources
and cosmology.
\end{abstract}

\end{frontmatter}

\section{Introduction}
A main theme of the conference was the multi-messenger approach
to the origin of cosmic rays.  The conference had a local
flavor that illustrates the strong Italian contributions to
major experimental activities in all three fields, gamma-ray astronomy, 
cosmic-ray experiments and neutrino astronomy.  In these brief
remarks I will point out some common techniques and approaches
and focus on a few important questions being addressed by
current experiments in particle astrophysics.  I make no attempt
to give a balanced and systematic review of the field here~\cite{ICRC2007}.

\section{Calibration}

One can distinguish two aspects of calibration.  One is to determine the response 
of a detector to a given beam or spectrum of particles.  The other is to
evaluate the level at which a source can be seen above background and make
sure the search algorithms would find a sufficiently strong source if it is present.
Computer simulations and test beams play a central role in both cases.  For example, the Large
Area Telescope (LAT) team on GLAST uses GEANT 4 to simulate the full detector
based on exposure of each component to accelerator beams.  The simulation is tuned
to data from further exposures of prototype assemblies to accelerator beams.  To address the
other aspect, a simulated sky has been created consisting of various likely sources
superimposed on the background of the Milky Way and an extra-galactic diffuse flux, 
as shown in Table~\ref{tab1}~\cite{Moiseev}.

\begin{table}
\centering
\caption{Contents of LAT data challenge sky~\cite{Moiseev}.}
\vspace{.3cm}
\begin{tabular}{l|c||l|c}
Galactic Sources & Number & ~Extra-galactic Sources & Number \\ \hline
Milky Way & 1 & Diffuse Extra-galactic & 1 \\ 
Moon & 1 & & \\ \hline
Pulsars & 414 & Bright variable AGN & 204 \\ \hline
Plerions & 7 & Faint steady AGN & 900 \\ \hline
SNR & 11 & GRB & 134 \\ \hline
XRB & 5 & GRB afterglow & 9 \\ \hline
OB associations & 4 & PBH & 1 \\ \hline
Small mol. cloud & 40 & Galaxy clusters & 4 \\ \hline
Dark Matter & $\sim$2 & Galaxies & 5 \\ \hline
`Other 3EG' & 120 & & \\ \hline
Solar flare & 1 & & \\ \hline
\end{tabular}
\label{tab1}
\end{table}

Pointing accuracy and angular resolution can be determined with data from a known source.
Gamma-ray telescopes typically use the Crab Nebula (e.g. VERITAS~\cite{Hanna})
as a calibration source.  The shadow of the Moon can also be used, particularly
for ground arrays operating at higher energy and with smaller statistics 
(e.g. ARGO-YBJ~\cite{dalistaiti}).  Angular resolution can also be determined
by measuring the same events with two different detectors.  An example of this is
a sub-array analysis with an air shower detector such as IceTop~\cite{Tilo}.
The ``checkerboard" analysis of the ARGO-YBJ carpet is another example.

\begin{figure}[t]
\centering
\includegraphics[width=\linewidth]{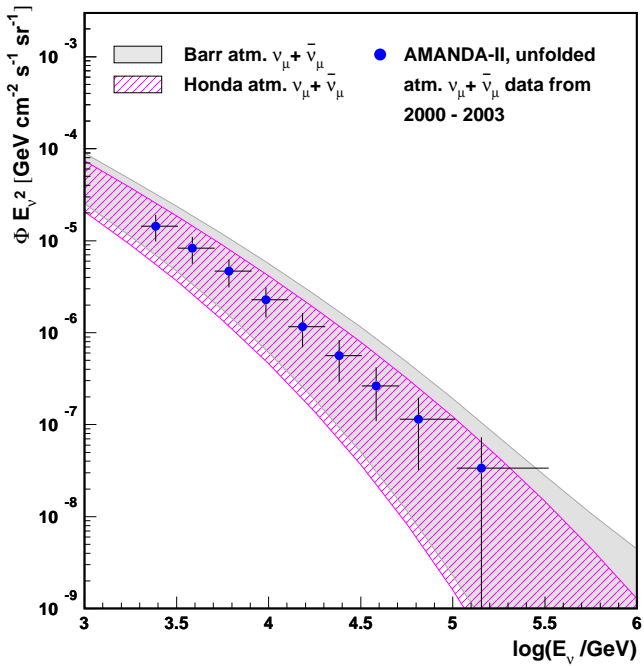}
\caption{Unfolded spectrum of atmospheric neutrinos by 
AMANDA-II~\cite{LeunemannKirsten}
compared to calculations of Refs.~\cite{Barretal,Honda2004}.  Shaded bands show
the range between the horizontal (upper border) and vertical flux (lower border).}
\label{Kirsten}
\end{figure}

For IceCube, the same principle can be used to check the pointing and angular
resolution of the neutrino telescope, to the extent that high-energy muons
in air shower cores tagged by IceTop are similar in the detector to
upward-moving neutrino-induced muons.  For air showers above the threshold
for IceTop, one can compare the direction reconstructed with the surface array with
the direction of the same event reconstructed independently by the deep array.
Since the surface array can be surveyed directly, this comparison checks pointing
as well as angular resolution.  Limitations are that showers above the threshold
for IceTop typically have several muons at the depth of the deep detectors of
IceCube and that coincident events that pass through both detectors are nearly
vertical.  To address the multiplicity problem, one can also use lower-energy events
that trigger only a single, inner IceTop station on the surface and compare the
line from that station to the center of the deep event with the direction
reconstructed by the IceCube reconstruction algorithm.  In her talk on
KM3NeT, the project to build a cubic kilometer neutrino detector in the
Mediterranean~\cite{deWolf}, 
Els de Wolf pointed out the possibility of deploying, perhaps 
temporarily, a floating air-shower detector above the deep-sea neutrino telescope.
Such a detector could have a variable spacing tuned to be able to reconstruct
small showers likely to produce a single muon that penetrates to the deep array.
It could also be moved to expand the range of angles explored.  

\begin{figure}[t]
\centering
\includegraphics[width=\linewidth]{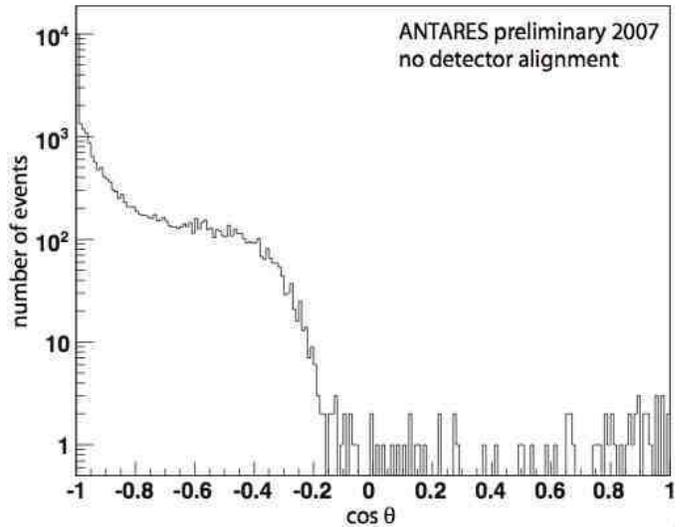}
\caption{Angular distribution of events in
Antares~\cite{Carr}.}
\label{Carr-fig}
\end{figure}

Apart from coincident events, the energy spectrum and angular distribution of
atmospheric muons and neutrinos are now rather well known in the TeV range.
Measuring and reconstructing both these distributions is an
important benchmark for neutrino telescopes.
The spectrum of diffuse neutrinos in AMANDA~\cite{deYoung} has been measured to approaching
100~TeV~\cite{LeunemannKirsten} and agrees reasonably well with expectation.  Some
atmospheric neutrinos have been identified with 9-string IceCube during 2006~\cite{Pretz}.
The first physics-quality data with IceCube is expected from the 22-string
detector currently operating in 2007 and in early 2008.  The next version of 
IceCube, including the new strings
of IceCube detectors to be deployed during the 2007/2008 season, is scheduled to start
a new run in April, 2008.  Antares has reported a preliminary measurement
of the zenith angle distribution of muons as shown in Fig.~\ref{Carr-fig}~\cite{Carr}.
Although there is not yet a comparison with the expected angular
distribution of atmospheric muons and neutrinos, the apparent emergence
of neutrino-induced muons slightly above the horizon is impressive.
If this interpretation of the plot is correct, it indicates good angular
resolution at the depth of 2050-2400 meters in the Mediterranean Sea.

The essential problem for calibration of large air shower arrays 
is that reconstruction of the
energy and mass of the  primary cosmic-rays they are designed to 
measure depends on extrapolation of properties of hadronic interactions
into regions inaccessible at accelerators.  Calibration of
the individual detectors on the ground is straightforward, either by
exposing them to accelerator beams or by using abundant GeV cosmic-ray
muons in situ.  The real problem is that reconstructing the
properties of the primary cosmic radiation from what is measured on 
the ground depends on comparison to simulations made with models
of hadronic interactions extrapolated orders of magnitude beyond
the regions of energy and phase space measured at accelerators.
Use of fluorescence detectors has the advantage that the measurement
of shower energy is more nearly calorimetric provided
complications of viewing angle, 
Cherenkov light background and variable atmospheric properties
can be overcome (for example with laser calibration shots).  But 
even in this case, there is a surprisingly large range of
predictions for the correction that has to be made for dark energy
in air showers (i.e. energy lost to neutrinos and energy carried into the
ground by muons).  In his talk on Auger,~\cite{Watson} Alan Watson showed a 6-7\%
contribution to the 24\% systematic uncertainty in the Auger flourescence
detector from this source.  

\section{Galactic sources}
Perhaps the most remarkable discovery in particle astrophysics of the last few
years is the large number of galactic sources observed in detail (including spatial
structure) by H.E.S.S. and described in this meeting in~\cite{Puehlhofer}.  The MILAGRO
experiment reports structure reflecting several sources in the Cygnus region as well
as a diffuse TeV gamma-ray background that is higher than expected from 
cosmic-ray propagation~\cite{Sinnis}.  Some of the H.E.S.S. sources are supernova 
remnants for which hadronic models seem likely~\cite{Voelk} in view of measurements of strong
magnetic fields, which make the electromagnetic interpretation more difficult,
in particular RX J1713-3946, in which the gamma-ray spectrum is observed to
100~TeV~\cite{J1713}.

The most direct signal of acceleration of protons as well as electrons in SNR would
be observation of neutrinos.  If the $>$TeV gamma-rays are decay products
of neutral pions produced in hadronic interactions ($p\rightarrow\pi^0\rightarrow \gamma+\gamma$),
then one can calculate from kinematics the corresponding spectrum of neutrinos from
$p\rightarrow\pi^+\rightarrow\mu^+ +\nu_\mu$.  This has been done for several H.E.S.S. TeV
gamma-ray sources in Ref.~\cite{Kappes}.  Optimistically, one expects to observe only a few
neutrino-induced muons per year in a cubic kilometer detector, and atmospheric 
backgrounds in the search window are comparable or somewhat larger.  In this situation,
strategies such as source stacking can be helpful.  Further references and discussion
are given in Ref.~\cite{Gaisser}.

In his talk on origin and acceleration of cosmic rays~\cite{Blasi}, Pasquale Blasi
emphasized the role of magnetic field amplification and non-linear effects at strong
shocks.  There is evidence from X-ray observations of young supernova remnants for
magnetic fields as high as 100~$\mu$Gauss, which allows acceleration of protons
to $>10^{15}$~eV.  Non-linear effects lead to differential spectra harder than $-2$
at the source.  The galactic spectrum with its observed differential index
of $-2.7$ must then be explained by a combination of several effects,
including propagation and time evolution
of the sources~\cite{Ptuskin}.  

Developing a full model of galactic cosmic rays
is a job that remains to be completed, but there is a growing consensus that
the knee of the cosmic-ray spectrum must be associated with
the beginning of the end of the spectrum of cosmic rays from sources in the Galaxy.
If so, the spectrum is expected to become increasingly dominated by heavy primaries
as the transition is approached.  In his talk on KASCADE~\cite{Jorg}, J\"{o}rg H\"{o}randel
showed evidence for an increase in the relative proportion of heavy nuclei with
increasing energy through the knee region.  This analysis also
brings up again the problem of how to extrapolate models of hadronic interactions
correctly.  The quantitative result of the analysis depends strongly on 
which event generator is used to interpret the KASCADE measurement of the ratio
of $\sim$GeV muons to electrons in the shower front.  Two event generators are
compared (SIBYLL 2.1 and QGSjet 01).  In both cases helium is apparently more
abundant above $10^{15}$~eV than protons.  However, the analysis with
SIBYLL 2.1 shows the CNO group as the most abundant component, whereas with QGSjet 01
helium is significantly more abundant that the other components.

\section{Extra-galactic sources}

A major recent result is the observation of a steepening of the ultra-high
energy cosmic ray spectrum above about $3-5\times 10^{19}$~eV by Hi-Res~\cite{Cao}
and Auger~\cite{Watson}.  One question that
arises is whether
the spectrum steepens because accelerators are reaching their maximum energy
or because of the effect of propagation and energy loss in the microwave background,
the Greisen-Kuz'min-Zatsepin (GZK) effect.  One way to confirm that it is the
GZK effect is to measure
the intensity of neutrinos in the EeV energy range.  Predictions for the 
intensity of GZK neutrinos that would be produced by $\sim10^{20}$~eV protons
during propagation depend on the cosmological evolution of the
sources and their energy spectra, 
as discussed here by Todor Stanev~\cite{Stanev}.  The question arises
how precisely and over what energy region it would be necessary to measure
the spectrum of GZK neutrinos in order to unfold information about the history
of cosmic-ray sources at large red shift and the source spectrum.  The expected
intensity of GZK neutrinos is such that a kilometer-scale detector might be
expected to detect one or two per year.  For this reason there is strong 
interest in using other techniques, such as radio~\cite{Besson} or 
acoustic~\cite{LeeThompson}, to achieve a
significantly larger effective detector volume.

The most frequently mentioned possibilities for the sources of ultra-high energy
cosmic rays are Gamma Ray Bursts and Active Galaxies.  In either case, depending on
the environment of the accelerator, a fraction of the energy could be lost in
interactions of accelerated hadrons in or near their sources.  If so, there
could be correlated hadronic production of both $>$TeV gamma-rays 
and neutrinos~\cite{Muecke,Waxman}.  

The recent AMANDA limit on
neutrino-induced muons from Northern Hemisphere GRBs~\cite{Kuehn} is close to
the benchmark Waxman-Bahcall prediction~\cite{Waxman}, and IceCube
is poised to extend the sensitivity significantly~\cite{Ignacio}.
With AGILE already in orbit~\cite{Tavani} and sensitive to GRBs on a range of
time scales, and particularly next year when GLAST begins operation,
there will be an increased number of tagged GRBs to define a
time window in which to look for neutrinos.  With time and direction windows
defined by an observed GRB, the background of atmospheric neutrinos is 
reduced to a very low level.

Leptonic models of TeV gamma-ray production in AGN are generally
favored over hadronic models, in which case the connection between the observed
intensity of TeV gamma-rays and possible neutrino fluxes would be lost.
A recent addition to the arguments against hadronic models of AGNs comes from
the observation by MAGIC of a flare from Mrk-501 in which the
 higher energy photons arrive a few minutes after lower energy 
ones~\cite{Bastieri}.  If the time delay reflects the time history of
the accelerated charged particles, it has a natural explanation in the context
of the electromagnetic, Synchrotron-Self-Compton Model; namely, that it may take longer
to accelerate electrons to higher energy, so the higher energy Compton up-scattered
photons would arrive later~\cite{MAGIC}. 

\section{Cosmology}

One motivation for the study of AGN spectra, apart from their intrinsic interest
as compact, energetic astrophysical sources, is that they have the potential
to probe the spectrum of extra-galactic background light in the infra-red
which reflects the history of star formation in early epochs
of the Universe~\cite{Stecker}.   The high-energy part of the spectrum
is cutoff in the TeV range by $\gamma\,\gamma\rightarrow e^+e^-$ for
sources with $z\sim 0.1$ and at lower energy for sources at larger redshift.
The possibility of such a systematic study was emphasized here as one motivating factor
in the proposal for a next generation gamma-ray telescope, CTA~\cite{Teshima}.
In his review of the subject~\cite{Stecker}, Stecker points out that cutoffs in
the range of tens of GeV are expected in the spectra of AGNs with $z\sim 2$,
a range accessible to GLAST.

Indirect detection of dark matter is a major goal of the satellite experiments
discussed at this conference.  The particle detectors 
(PAMELA~\cite{Papini}, AMS~\cite{Battiston}) will be searching
for an excess of anti-matter over what is expected from cosmic-ray propagation
through the interstellar medium.  Such excesses would be a natural expectation, for
example from WIMP pair annihilation in concentrations of dark matter, because
particles and antiparticles would be produced in equal abundance.  As emphasized
in Roberto Battiston's talk~\cite{Battiston}, not only positrons and anti-protons
could be observed, but also anti-deuterons.  The latter would be a particularly
clean signal of annihilation~\cite{Baer}
because of the extreme difficulty of producing an anti-deuteron
at low energy in the collision of a cosmic ray with an interstellar nucleus
at rest.  Because of the very high energy threshold for the process,
any $\bar{d}$ produced on a stationary target would be quite energetic.

Gamma-rays are also a potential signature of dark matter, particularly
in the case where the signal produces a peak or line in the spectrum.
The search for a signal of dark matter is among the principal 
goals of GLAST~\cite{Moiseev}.  In his talk here, 
Lars Bergstrom~\cite{Bergstrom} described the line signature
from WIMP pair annihilation into $\gamma\,\gamma$ with $E_\gamma = M_{\rm WIMP}$.
The branching ratio for this mode is likely to be small, but the
signature is distinctive.  In general, any gamma-ray product of dark matter
interactions should point to regions of high mass concentration in the Galaxy.

Finally, I conclude with a reminder that the particle detectors PAMELA and
AMS will make measurements of spectra of galactic cosmic rays with unprecedented precision.
They will also map the effects of the Sun on cosmic rays and observe
solar energetic particles.  A nice example shown here was the spectrum of 
the December 13, 2006 solar flare observed with PAMELA~\cite{Papini}.

\section*{Acknowledgments}
This work is supported in part by the U.S. 
Department of Energy under DE-FG02 91ER 40626.


\end{document}